\documentclass{article}

\usepackage{arxiv}

\usepackage[utf8]{inputenc} 
\usepackage[T1]{fontenc}    
\usepackage{hyperref}       
\usepackage{url}            
\usepackage{booktabs}       
\usepackage{amsfonts}       
\usepackage{nicefrac}       
\usepackage{microtype}      
\usepackage{graphicx}
\usepackage{natbib}
\usepackage{doi}
\usepackage{color}
\usepackage{tabularray}
\usepackage{float}

\title{Using causal inference to avoid fallouts in data-driven parametric analysis: a case study in the architecture, engineering, and construction industry}

\author{ {\hspace{1mm}Xia Chen}$^*$\\
            Sustainable Building Systems \\
            Leibniz University Hannover \\
            Hannover, Germany \\
	\texttt{xia.chen@iek.uni-hannover.de} \\
	\And
 {\hspace{1mm}Ruiji Sun}\\
            Center for the Built Environment \\
            University of California \\
            Berkeley, USA \\
	\texttt{ruijis@berkeley.edu} \\
 \And
  {\hspace{1mm}Ueli Saluz}\\
            Sustainable Building Systems \\
            Leibniz University Hannover \\
            Hannover, Germany \\
	\texttt{ueli.saluz@iek.uni-hannover.de} \\
 \And
 {\hspace{1mm}Stefano Schiavon}\\
            Center for the Built Environment \\
            University of California \\
            Berkeley, USA \\
	\texttt{schiavon@berkeley.edu} \\
 \And 
 {\hspace{1mm}Philipp Geyer}\\
            Sustainable Building Systems \\
            Leibniz University Hannover \\
            Hannover, Germany \\
	\texttt{philipp.geyer@iek.uni-hannover.de} \\
}

\renewcommand{\shorttitle}{Using causal inference to avoid fallouts in data-driven parametric analysis}

\begin{document}
\maketitle

\begin{abstract}
	The decision-making process in real-world implementations has been affected by a growing reliance on data-driven models. We investigated the synergetic pattern between the data-driven methods, empirical domain knowledge, and first-principles simulations. We showed the potential risk of biased results when using data-driven models without causal analysis. Using a case study assessing the implication of several design solutions on the energy consumption of a building, we proved the necessity of causal analysis during the data-driven modeling process. We concluded that: (a) Data-driven models' accuracy assessment or domain knowledge screening may not rule out biased and spurious results; (b) Data-driven models' feature selection should involve careful consideration of causal relationships, especially colliders; (c) Causal analysis results can be used as an aid to first-principles simulation design and parameter checking to avoid cognitive biases. We proved the benefits of causal analysis when applied to data-driven models in building engineering.
\end{abstract}


\section{Introduction}
In recent decades, successful implementations of machine learning (ML) methods, with the momentum of growing data volume, have brought the data-driven approach into various engineering domains. Together with empirical domain knowledge analysis and first-principles simulations, ML methods have become a handy tool for both academic research and industrial application \citep{bertolini2021machine,lecun2015deep,raschka2020machine}. Due to their end-to-end learning behavior, good generalization performance, and fast prediction response, they are favored by researchers and engineers, and are gradually being integrated as a decision-making or analysis assistance tool in the architecture, engineering, and construction (AEC) industry \citep{dimiduk2018perspectives,marcher2020decision,seyedzadeh2018machine}.

The advantage of MLs’ wide adaptability comes from their ability to directly capture hidden patterns from the data during training by minimizing the error, instead of explicitly modeling the physical process with domain knowledge context. However, their prerequisite in modeling processes for a proper performance need to assume that all input variables are independent, or even \textit{independent and identically distributed} (i.i.d.) \citep{scholkopf2022causality} by default. That is, the probability distribution of each value (variable) should have no dependence on other values. However, in reality, especially in engineering domains, a case usually requires considering different factors in an interdisciplinary manner. For instance, during the building design or construction phase, the objectives commonly involve building energy performance, environmental impact, cost, occupant’s comfort, etc., simultaneously. The well-known mantra in statistics: "\textit{Correlation does not imply causation}" \citep{aldrich1995correlations,pearl2018book}, is not sufficiently considered in engineering scenarios \citep{chakraborty2019advanced,hegde2020applications} when ML methods are used. Unlike first-principles simulations, which encode causal relationships between variables in explicit physical equations, data-driven processes do not include this information. Lacking this process understanding might lead to false implementation and reliability issues for engineers and domain experts. This false implementation situation raises the risk of biased results and spurious conclusions because ML methods rely heavily on the information carried from the distribution of observed data and large predefined sets \citep{scholkopf2021toward}. 

In this study, we propose a synergetic framework. This framework integrates empirical domain knowledge from human experts, simulations, and data-driven methods. Our aim is to promote their combined use in general engineering analysis. We employ a real-world building engineering scenario in the design phase. In this scenario, we highlight a potential "fallout" situation that could arise in a data-driven modeling analysis, followed by the introduction of the causal analysis process. We show the need for causal dependencies checks among variables during the data-driven process for two main reasons. First, fitting data through data-driven methods without considering causal dependencies carries potentially biased estimates. They result in spurious conclusions and risks in engineering scenarios. These limitations are present regardless of the type of machine learning methods, and cannot be eliminated via model accuracy improvement. Secondly, in engineering scenario analysis, the discovery of causal dependencies and the construction of a causal skeleton are practical tools. They help to cross-validate data with domain knowledge, examine whether potential cognitive biases exist in the simulation process, and aid in knowledge discovery. We believe these tools create a crucial link between data-driven methods and human reasoning in design and engineering processes.

\section{Framework and methodologies}
\subsection{Synergetic Framework between Experience, Simulation, and Data-driven Methods}
In engineering, the tools we use for modeling and decision-making can be classified into three main categories: empirical domain knowledge, first-principles simulation, and data-driven models:
\begin{itemize}
    \item \textbf{Empirical domain knowledge} is a carrier of individual and past professional experience, providing a fundamental drive to understand, interact, and make decisions in a system. This includes heuristic rules or "rules of thumb" – quick, intuitive information set. However, it is limited by personal competence and often lacks reproducibility.
    \item \textbf{First-principles simulation} is a process based on abstract symbolic abstraction,  using mathematical equations and physical/chemical laws to govern the behavior of a system. By starting from basic principles and building up to an understanding of complex phenomena, first-principles simulations are also referred to as "white-box models".
    \item \textbf{Data-driven method} is a computational process based on available data rather than theoretical principles or physical laws. These processes employ ML algorithms, statistical models, and data analysis techniques to extract patterns and relationships from datasets. These patterns are then used to make predictions or generate insights about the system, functioning as "black-box models".
\end{itemize}
Table \ref{tab:tab1} illustrates the main advantages and disadvantages of these three major categories we rely on in engineering.

\definecolor{Black}{rgb}{0,0,0}
\begin{table}
	\caption{The characteristics of relying on empirical domain knowledge, first-principles simulation, and data-driven approach for engineering modeling.}
\centering
\begin{tblr}{
  width = \linewidth,
  colspec = {Q[158]Q[404]Q[379]},
  cell{1}{2} = {c},
  cell{1}{3} = {c},
  hlines,
  hline{2,5} = {-}{black},
  hline{3-4} = {-}{Black},
}
                                                            & \textbf{Advantages}                                                                                                                                                                                                                                               & \textbf{Disadvantages}                                                                                                                                                                                                      \\
{\textbf{Empirical }\\\textbf{domain }\\\textbf{knowledge}} & {\labelitemi\hspace{\dimexpr\labelsep+0.5\tabcolsep}No
  extra efforts needed for modeling;
  \\\labelitemi\hspace{\dimexpr\labelsep+0.5\tabcolsep}Foundation
  for scientific inquiry and hypothesis testing;
  }                                                & {\labelitemi\hspace{\dimexpr\labelsep+0.5\tabcolsep}Rule
  of thumb, heavily relies on personal ability;
  \\\labelitemi\hspace{\dimexpr\labelsep+0.5\tabcolsep}Limited
  extent and reliability in non-standard cases;
  } \\
{\textbf{First-principles }\\\textbf{simulation}}           & {\labelitemi\hspace{\dimexpr\labelsep+0.5\tabcolsep}Good
  interpretability;
  \\\labelitemi\hspace{\dimexpr\labelsep+0.5\tabcolsep}Flexible
  in modeling details;
  \\\labelitemi\hspace{\dimexpr\labelsep+0.5\tabcolsep}Large
  amount of output variable;
  } & {\labelitemi\hspace{\dimexpr\labelsep+0.5\tabcolsep}Time-consuming
  in detailed simulation; 
  \\\labelitemi\hspace{\dimexpr\labelsep+0.5\tabcolsep}Modeling
  efforts required in each new scenario;
  }                  \\
\textbf{Data-driven method}                                 & {\labelitemi\hspace{\dimexpr\labelsep+0.5\tabcolsep}Fast
  response in prediction;
  \\\labelitemi\hspace{\dimexpr\labelsep+0.5\tabcolsep}Universal approximator;
  \\\labelitemi\hspace{\dimexpr\labelsep+0.5\tabcolsep}End-to-end
  learning behavior;
  }      & {\labelitemi\hspace{\dimexpr\labelsep+0.5\tabcolsep}Black-box,
  trustfulness issues;
  \\\labelitemi\hspace{\dimexpr\labelsep+0.5\tabcolsep}Data-hungry for training;
  \\~
  }                                            
\end{tblr}
\label{tab:tab1}
\end{table}

\begin{figure}[h]
	\centering
	\includegraphics[width=16cm]{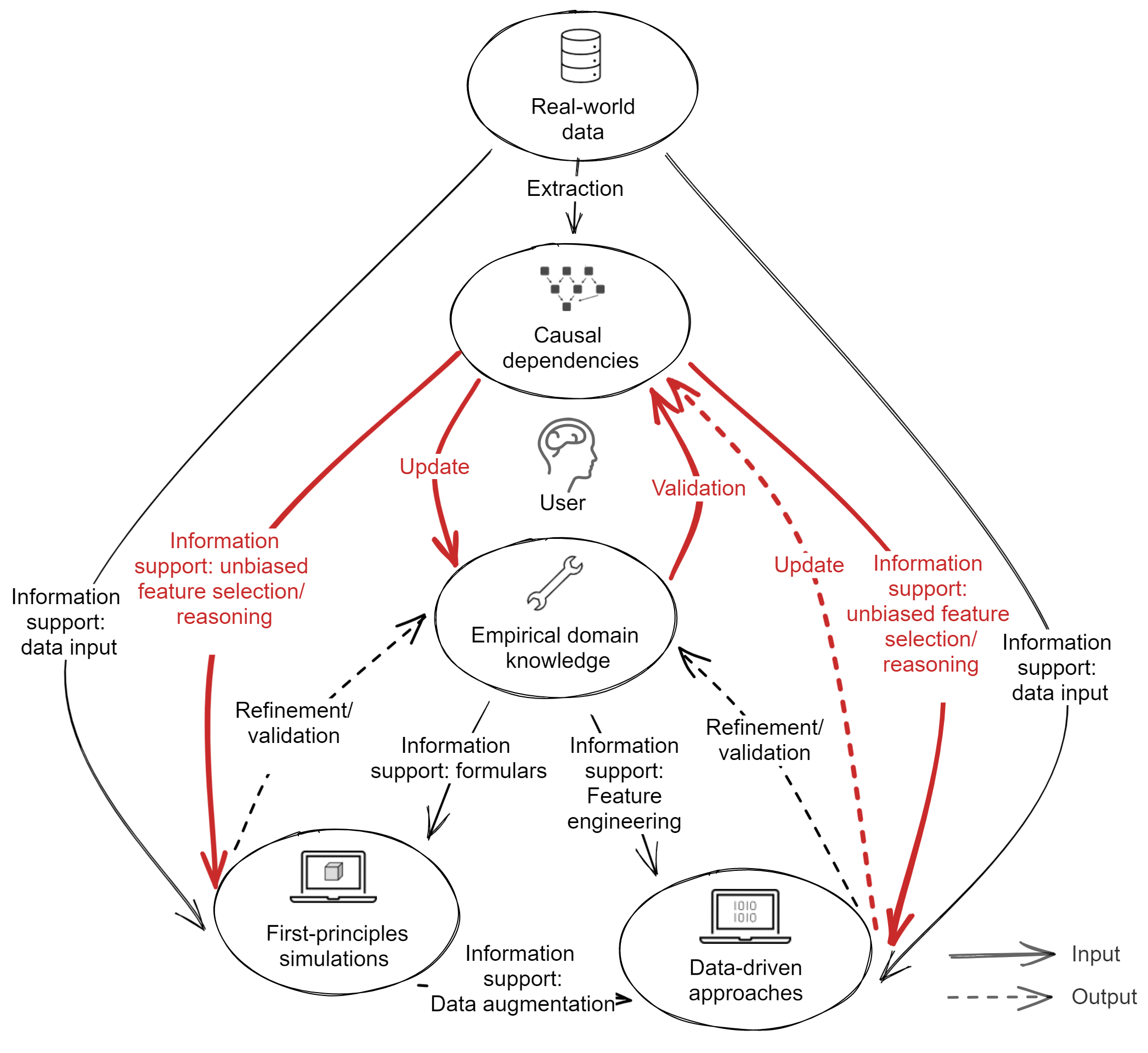}
	\caption{Illustration of the potentially synergetic nature of the three main engineering modeling processes. Causal dependencies extracted from data represent a type of invariant, transferable knowledge, which play a vital role in offering a feedback loop and interacting with the user’s empirical domain knowledge. Except for the data and domain knowledge input, the causal dependencies information support contributes to validation for first-principles simulation, and unbiased estimation/reasoning for data-driven methods. The red arrows indicate how the causal relationships interact with other engineering modeling approaches.}
	\label{fig:1.png}
\end{figure}

In engineering scenarios, we possess, reuse, and iterate on invariant patterns that can be applied to many cases. These patterns form what is known as knowledge and experience \citep{chen2022introducing}. For instance: the case of sinking library \footnote{The sinking library: A famed college library is gradually sinking into the ground because its architect failed to take the weight of the books into account.} updates our consideration of the relationship between building type/usage and building structural engineering. In first-principles simulations, the relationships between these variables are naturally embedded into symbolic formulas and numerical modeling processes as knowledge. However, this type of information input is absent in the data-driven process. We propose that the data-driven method should include an additional, transferable piece of information: causal dependencies among variables. We illustrate this idea in Figure \ref{fig:1.png}, demonstrating how causal dependencies extracted from the data interact with experiential domain knowledge, first-principles simulations, and data-driven approaches in a synergetic manner.

In Figure \ref{fig:1.png}, red arrows indicate how causal relationships interact with other engineering modeling approaches. Causality is commonly confused with correlation, but the former presents a different interpretation from observational data: It analyzes the asymmetric change and response between cause and effect, aids in analyzing interventional scenarios, counterfactuals, and answers "what-if" questions. This reasoning ability is essential for informative and sequential decision-making support. Additionally, the extracted causality information provides a feedback loop for users to validate and update their domain knowledge, fostering unbiased modeling.

\subsection{Causality}
Causality research has become a critical topic and has made substantial contributions across various fields with the widespread adoption of data-driven methods in the past decade \citep{scholkopf2022causality,spirtes2010introduction}. Causal inference examines parameters or properties, considering cause-effect logical sequences to avoid unrealistic conclusions. For a systematic discussion of causal inference research, we refer to research to the works of Pearl \citep{pearl2009causal}, Spirtes et al. \citep{spirtes2010introduction,spirtes2000causation}, and Peters et al. \citep{peters2017elements}.

Our previous research \citep{chen2022introducing} introduced causal inference into the energy-efficient building design process, using a four-step framework that combined causal structure finding and causal effect estimation. In this study, we aim to demonstrate the importance of checking causal dependencies in the context of the general AEC domain. This section briefly clarifies foundational ideas related to causal analysis.

\textbf{Causal finding algorithms} are methods for identifying and returning equivalence classes of proper causal structure based on observational data in an unsupervised, data-driven manner. Essentially, they distinguish asymmetries in sampling distributions to identify feature dependencies and causal directions. Typical causal structure finding algorithms based on observational data fall into three categories: constraint-based, score-based, and hybrid \citep{kalisch2014causal}. In this study, we chose one of the typical score-based methods with a greedy mechanism \citep{devore1996some}, Greedy-Equivalent-Search (GES) \citep{chickering2002learning,chickering2002optimal}. 

\begin{figure}[h]
	\centering
	\includegraphics[width=16cm]{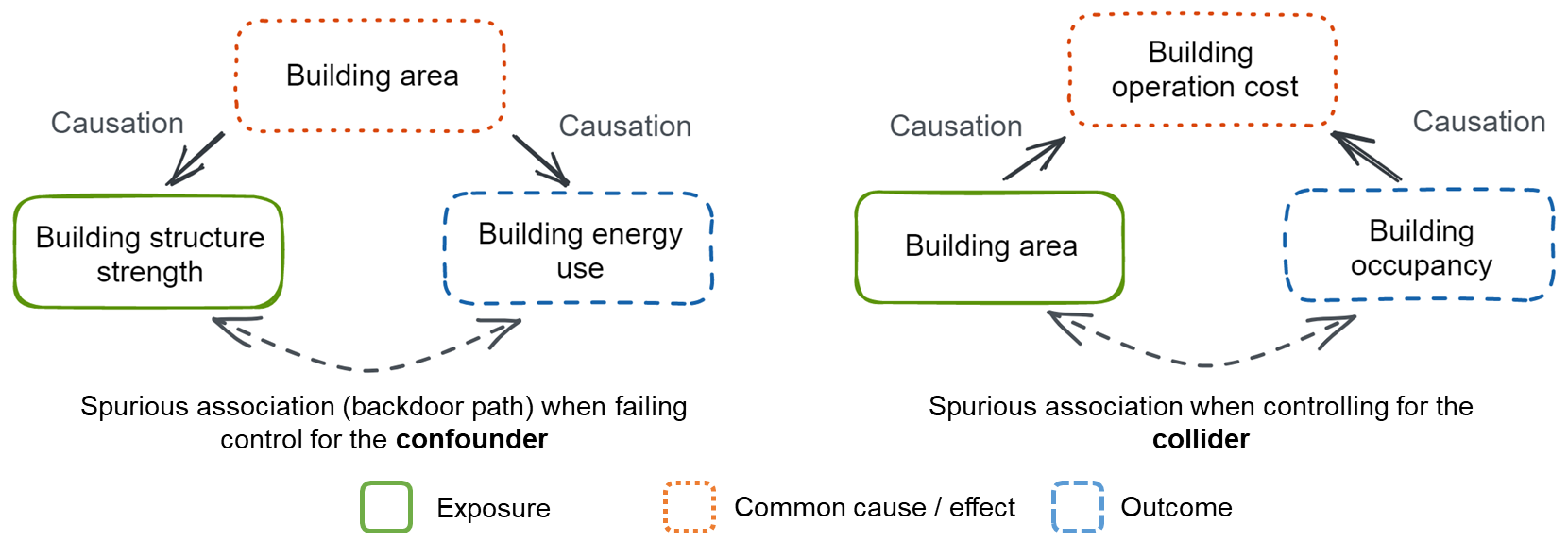}
	\caption{Causal confounder and collider examples in the context of architectural engineering domain. Failing to identify the causal relationship cause spurious association (backdoor path) and biased results. Left: confounder bias when the common cause is not controlled; Right: collider bias when the common effect is controlled.}
	\label{fig:2.png}
\end{figure}

\textbf{Directed Acyclic Graphs (DAGs)} are graph diagrams composed of variables (nodes) connected via unidirectional arrows (paths) to depict hypothesized causal relationships \citep{judea2010introduction}. A causal skeleton DAG with a fixed structure embeds the causal dependencies of given data. A DAG demonstration in the building engineering domain is presented in Figure 2. Three major types of DAG structure combinations are:
 \begin{itemize}
     \item \textbf{Directed path} denotes a directed edge \textit{x→ y} of \textit{x} (cause) on \textit{y} (effect). Intuitively, it means that \textit{y} is directly influenced by the status of \textit{x}, altering \textit{x} by external intervention would also alter \textit{y}.
     \item \textbf{Backdoor path} exists in two variables in a confounding structure where the common cause is not controlled (Figure \ref{fig:2.png}, left), or two variables in a collider structure where the common effect is controlled, effect variables connected by this backdoor path have a non-causal association and would lead to potential bias with distorting association.
     \item \textbf{Closed path} exists in collider structures where two variables have the same effect (Figure \ref{fig:2.png}, right). Unlike directed and backdoor paths, this path is causal-wise irrelevant: there is no causal path between the two variables via the collider structure, unless the common effect is controlled.
 \end{itemize}

\textbf{DAG rules} are principled structural guidelines that enable users to investigate cases for identifying appropriate sets of covariates in complex DAGs and for removing structural bias through adjustments, e.g., d-Separation, backdoor criterion \citep{pearl2000models}, and their extensions.

DAGs, often defined by prior knowledge, could be incomplete \citep{guo2020survey}. In the development of a causal diagram, users utilize their best available prior knowledge to set up the most plausible causal diagram. Subsequently, they adhere to strict DAG rules to identify the causal dependencies between given exposure inputs and the target outcome from the case. In the remaining content, all DAGs are generated and modified by DAGitty \citep{textor2015drawing,textor2016robust}.

A "fallout" situation in this context refers to an instance in causal analysis where an indirect, biased relationship exists between an exposure (or cause) and an outcome (or effect), primarily due to the presence of the backdoor path or the opening of the closing path.

\subsection{Machine Learning}
In this study, we focus on ML methods applied to supervised learning tasks, which typically involve addressing a classification or regression problem with labeled data. To ensure that the fallout is irrelevant to the type of data-driven methods used, we examined three mainstream ML methodologies \citep{singh2016review}: tree-based models \citep{clark2017tree}, kernel machines \citep{hofmann2008kernel}, and neural networks \citep{lecun2015deep}, which are mechanistically different and widely applied in engineering domains \citep{seyedzadeh2018machine,chakraborty2019advanced,hegde2020applications}. Brief introductions to their mechanisms are given in Appendix. Beyond these methods, the evaluation of uncertainties is critical for supporting the decision-making process \citep{chen2022machine,tian2018review}, leading us to include a probabilistic, tree-based, gradient-boosting surrogate model – NGBoost \citep{duan2020ngboost} in our case study. Instead of generating output as a point prediction, the design of NGBoost incorporates a predictive uncertainties quantification process, offering insights into the output range within the set of feature input descriptions in a data-driven manner.

\section{Case study}
\subsection{Scenario Setup}
We studied the effect of different designs on energy use for heating (Energy Usage Intensity of heating, \textit{EUI Heating}) by varying \textit{insulation standards} and \textit{heating systems}.

To prepare our dataset, we utilized a parametric office building simulation model. This model represents a realistic design space by incorporating a wide range of configurations for building components and zones to train our ML models (training data). The causal reasoning within space is validated by a real-world design project from our previous research \citep{Chen2022} (test case): a mixed-usage, four-floor building known as Building.Lab, located on a tech campus in Regensburg, Germany. 

We simulated three sets of thermal characteristics to explore design variations in insulation values. These were based on existing standards: the 2020 German Energy Act for Buildings (\textit{GEG}), Net Zero Energy Building (\textit{NZEB}), and \textit{Passive House}. These standards, from baseline to high, have different requirements for components’ thermal conductivity (U-values), with a higher standard indicating better building thermal behavior and less energy loss. We also configured three typical building heating systems: \textit{boiler}, air-sourced heat pump (\textit{ASHP}), and district heating (\textit{DH}). For the modeling tool, we used Grasshopper \citep{mcneel2022grasshopper}, with Honeybee \citep{ladybug2021ladybug} serving as a high-level simulation interface for EnergyPlus.

In terms of data-driven modeling approaches, as discussed in Section 2.3, we applied Decision Tree (DT), Support Vector Machine for Regression (SVR), Artificial Neural Network (ANN, with Multi-Layer Perception chosen as a basic variation), and NGBoost across all scenarios.  

We applied three metrics to facilitate performance comparison across different numerical scales of results: Normalized Root Mean Square Error (NRMSE), Symmetric Mean Absolute Percentage Error (SMAPE), and Coefficient of determination (R-squared or $R^2$). We chose $R^2$ as our primary reference. The reasoning behind this choice and detailed interpretations of these three metrics are available in \citep{chicco2021coefficient}. 

Table \ref{tab:tab2} lists the input features from the simulation, their ranges, and the corresponding test case setting. To avoid the extrapolation problem (which arises when the test case sample falls outside of the given training dataset's convex hull \citep{balestriero2021learning}), all feature values in the test case are within the range of training data. We fitted and fine-tuned ML models with the training data to achieve well-generalization performance, and used them later to predict different scenarios in the test case, in which all values are extracted from the Building.Lab project in a real-world context. 

\definecolor{Gray}{rgb}{0.498,0.498,0.498}
\begin{table}
	\caption{Ranges in training data features and value extracted from the test case. All values in the test case are extracted from the Building.Lab project for the case study.}
\centering
\begin{tblr}{
  width = \linewidth,
  colspec = {Q[517]Q[229]Q[194]},
  column{2} = {c},
  column{3} = {c},
  hline{1} = {-}{},
  hline{2,18} = {-}{Gray},
}
\textbf{Building feature / Variable}                                      & \textbf{Training data range}                           & \textbf{Test case setting} \\
\textit{Orientation [}°\textit{]}                                         & {[}0, 180]                                             & 12.5                       \\
\textit{Number of Floors }                                                & {[}1, 10]                                              & 4                          \\
\textit{Floor Height [m]}                                                 & {[}2.8, 4.5]                                           & 3.48                       \\
\textit{Open Office: Heating Setpoint [}°C\textit{]}                      & {[}21, 24]                                             & 22                         \\
\textit{Open Office: Air Change Rate (ACH) [1/h]}                         & {[}4, 6]                                               & 4                          \\
\textit{Open Office: People Per Area (PPA) [people/m²]}                   & {[}0.05, 0.2]                                          & 0.15                       \\
\textit{Volume [m³]}                                                      & {[}4400, 146000]                                       & 6807                       \\
\textit{Area\textsuperscript{1} {[}m²]}                                   & {[}1300, 36000]                                        & 1956                       \\
\textit{Construction Area\textsuperscript{2} {[}\%]}                      & {[}3, 11.5]                                            & 6                          \\
\textit{Window to Wall Ratio North [\%]}                                  & {[}0, 0.7]                                             & 0.5                        \\
\textit{Window to Wall Ratio East [\%]}                                   & {[}0, 0.7]                                             & 0.45                       \\
\textit{Window to Wall Ratio South [\%]}                                  & {[}0, 0.7]                                             & 0.34                       \\
\textit{Window to Wall Ratio West [\%]}                                   & {[}0, 0.7]                                             & 0.23                       \\
\textit{Insultation Standard}                                             & base, medium, high                                     & Unknown                    \\
\textit{Heating System}                                                   & Boiler, ASHP\textsuperscript{3}, DH\textsuperscript{4} & Unknown                    \\
\textit{Energy Usage Intensity (EUI) Heating [kWh/m\textsuperscript{2}a]} & {[}14.6, 327.1]                                        & Unknown                    
\end{tblr}
\textit{\textsuperscript{1} Floor area gross; \textsuperscript{2} Areas covered by walls, columns, or any structural elements; \\\textsuperscript{3} ASHP: air-sourced heat pump; \textsuperscript{4} DH: district heating;}
\label{tab:tab2}
\end{table}

Further information regarding modeling configuration, data generation process, and training strategy of data-driven models are available in Appendix. 

With the set training data and test case, we first set up two scenarios: 
\begin{itemize}
    \item \textit{\textbf{Scenario I}}: Full-scale modeling with all input features for EUI heating prediction as the benchmark.
    \item \textit{\textbf{Scenario II}}: Masked input features, which represent common situations in real-world engineering scenarios - feature selection by domain knowledge, or only some features are observable/available during data collection.
\end{itemize}
Scenario I presents an ideal case in research or engineering, demonstrating how the data-driven process helps to provide analytical insights into potential design scenarios. However, in real-world cases, data is rarely as complete as in an ideal scenario due to the presence of unobserved factors, the need for simplification because of the expensive data collection and computation efforts, or subjective manual filtering by end-users using their own domain knowledge or analytical tools. In Scenario II, we illustrate the potential risks of introducing subjective bias associated with such incomplete data: We selected the following input features that are typically cared for by architects or engineers in the building design phase for energy performance evaluation \citep{marcher2020decision,chen2022hybrid,roman2020application}: \textit{Open Office: Heating Setpoint, Open Office: ACH, Open Office: PPA, Volume, Area, }and\textit{ Window to Wall Ratios}.

In both scenarios, ML models are fitted and evaluated using the training data, then used to predict the output with test case inputs plus different insulation standard and heating system combinations. 

\subsection{Benchmark and Fallout}
Table \ref{tab:tab3} presents the prediction results of different models fitted with the training data in the setting of both scenarios. The results demonstrate the model capabilities in this training case; all ML methods trained by full input features show acceptable performance. The $R^2$ of all models is above 0.85, while ANN and NGBoost reach an accuracy above 0.95. With the masked feature setting but the same training process as in Scenario I, the result shows only a minor performance decrease in Scenario II: All models maintain their accuracy ($R^2$) above 0.8, with ANN and NGBoost remaining around 0.9. We even observed a slight performance improvement for SVR in Scenario II. NRMSE and SMAPE results also align with this interpretation (see Appendix).

\begin{table}
	\caption{5-fold cross-validation performance result comparison of different models: Scenario I \& II}
\centering
\begin{tblr}{
  width = \linewidth,
  column{1} = {c},
  column{2} = {c},
  column{3} = {c},
  hline{1-2,6} = {-}{},
}
\textbf{Model}         & \textbf{R\textsuperscript{2 }(Scenario I)\textsuperscript{}} & \textbf{R\textsuperscript{2} (Scenario II)} \\
\textit{Decision Tree} & 0.86                                                         & 0.81                                        \\
\textit{SVR}           & 0.87                                                         & 0.87                                        \\
\textit{ANN}           & 0.96                                                         & 0.94                                        \\
\textit{NGBoost}       & 0.95                                                         & 0.88                                        
\end{tblr}
\label{tab:tab3}
\end{table}

\begin{figure}[h]
	\centering
	\includegraphics[width=14cm]{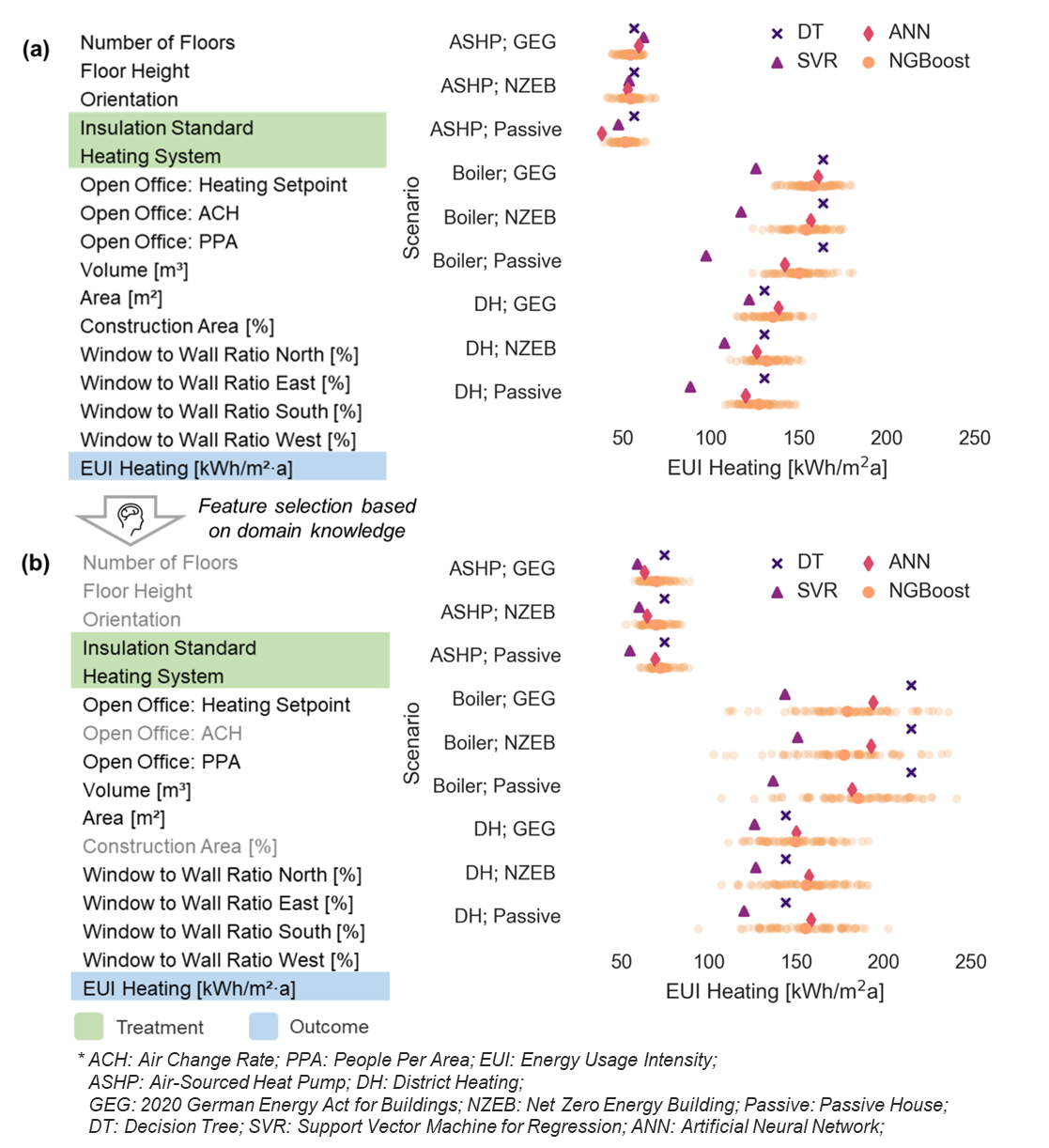}
	\caption{Test case prediction result based on: (a) Scenario I trained with full-scale features; (b) Scenario II trained with masked features selected manually based on domain knowledge. In both subgraphs, the left part shows the selected features with set exposures (treatment inputs we want to vary) and outcome based on the scenario, while the right part is the prediction result on the test case: The y-axis lists different combinations of insulation standard and heating system setting, while the x-axis gives the EUI Heating prediction result from different models (by different markers).}
	\label{fig:3.png}
\end{figure}

Next, the test case is fed with variations for insulation standard and energy system into trained models for both scenarios. We illustrate the corresponding results from different variation combinations in Figure \ref{fig:3.png}. 

Based on the result of Scenario I (Figure \ref{fig:3.png}a, right), we concluded the following insights:
\begin{enumerate}
    \item The test case prediction results from ANN and NGBoost are more similar; they also achieve better accuracy in the training process evaluation.
    \item The choice of the energy system is the factor that affects the EUI heating the most, with the air-source heat pump (ASHP) system requiring the least energy consumption, and the boiler system the most.
    \item Regardless of heating system variation, higher building component thermal standards contribute to reducing total energy consumption, as expected.
\end{enumerate}

With almost the same accuracy performance, the test case prediction result in Scenario II displays unusual patterns that contradict domain intuition, as shown in Figure \ref{fig:3.png}b. Although the choice of the heating system still shows the deterministic impact on EUI heating, the trend acts oppositely in insulation standard variation: The difference between the building insulation standards is either barely noticeable or even presents an inversed trend. Within the same heating system choice, a higher insulation standard results in more energy consumption in heating. This opposing trend even shows in the ANN, which achieves 0.94 in $R^2$ during performance evaluation. Furthermore, we observed a drastic increase in the uncertainty range in the output of NGBoost compared to Scenario I (see orange scatter distributions in Figure \ref{fig:3.png}).

Based on the result from Scenario II, \textbf{wrong conclusions} could easily be drawn, potentially misguiding decision-making process in real-world projects or research, e.g.: 

\textit{“In this case, insulation standard choices are unimportant, or adapting a lower insulation standard could help to reduce the energy usage of the building.”}

This conclusion drawn from Scenario II clearly conflicts with the result from Scenario I and with common knowledge. We refer to Scenario II as a case of biased estimation or a fallout. This fallout is directly linked to potential economic and energy loss, as well as risks if implemented in real-world engineering construction scenarios. Given that the cost of implementing higher insulation standards in buildings is typically an important factor, this misleading conclusion could lead to the decision of investment reduction or underestimation.

Such uncertain performance in the analysis could cause severe trust issues when adopting data-driven methods in engineering scenarios and decision-making processes. This is because real-world scenarios are less likely to provide complete data without hidden variables. It is less relevant to the modeling approach and cannot be ruled out by performance evaluation. As the only difference between the two scenarios is the feature selection, a closer examination of the input analysis, more specifically, the causal dependency analysis, is necessary.

\subsection{Causal Dependencies Analysis}
\begin{figure}[H]
	\centering
	\includegraphics[width=16.5cm]{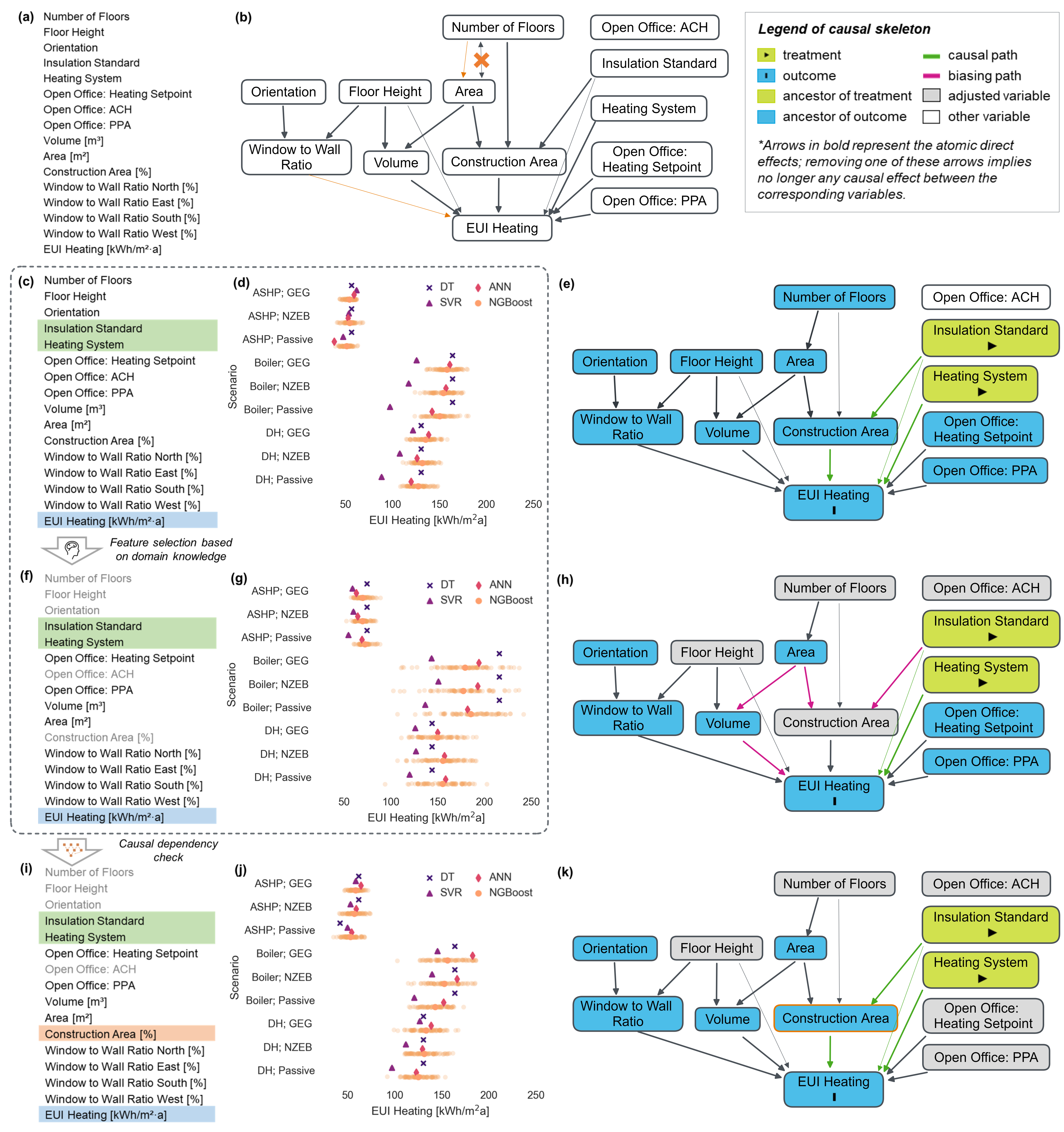}
	\caption{Causal dependencies analysis process, the dotted box is the content of Figure \ref{fig:3.png}. (a), (b): Causal structure finding via GES: knowledge extraction based on the training dataset. Minor skeleton adjustments via domain knowledge are marked in orange; (c), (d), and (e): Scenario I; (f), (g) and (h): Scenario II: Blocking Construction Area would close the direct causal path from \textit{Insulation Standard $\rightarrow$ Construction Area $\rightarrow$ EUI Heating}, and open a biasing path from \textit{Insulation Standard $\rightarrow$ Area $\rightarrow$ Volume $\rightarrow$ EUI Heating}, which leads to spurious conclusion; (i), (j) and (k): Corrected Scenario II with no biasing path.}
	\label{fig:4.png}
\end{figure}

From a causal inference analysis perspective, the hidden relationships among input features cause the biased outcomes observed in Scenario II. Similar cases have been discussed in medical statistic research \citep{patil1981causal}. In this section, we demonstrate that for the AEC domain, causal discovery can aid designers and engineers in comprehensively examining whether hidden relationships have been neglected and, by controlling them accordingly, avoid subjective bias and biased estimation. For a more intuitive engineering interpretation and evaluation, we expand upon Figure \ref{fig:3.png} and present a coherent causal dependencies analysis process to demonstrate that the analysis help avoid the fallout situation, as shown in Figure \ref{fig:4.png}. 

The first step of causal dependencies analysis is causal discovery, which is responsible for extracting a causal skeleton from training data in an unsupervised manner. The skeleton and process itself bring a critical nexus for connecting data-driven results with domain knowledge validation through causal skeleton pruning. In our case study, the pruning process is relatively straightforward, as demonstrated in Figure \ref{fig:4.png}b; only minor adjustments (marked in orange) are made based on the original skeleton generated by GES: 
\begin{enumerate}
    \item Adding a causal dependency (arrow) from \textit{Window to Wall Ratio (WWR)} to \textit{EUI Heating}, since the causal connection between these two variables is slightly indirect. This is due to us manually merging all WWRs into one for a more simplified illustration.
    \item Replacing the bidirectional arrow between \textit{Number of Floors} and \textit{Area} with a unidirectional arrow, as the number of floors is typically a variable given based on urban regulations determining the feasible floor area on a specific site.
\end{enumerate}
Subsequent to the setup of the causal skeleton, the exposure inputs \textit{(Insulation Standard} and \textit{Heating System}) and the target outcome (\textit{EUI Heating}) are integrated into the skeleton,  thereby establishing the causal flow, as illustrated in Figure \ref{fig:4.png}e. Based on the skeleton and scenario setting, we identified three crucial intermediate features: \textit{Window to Wall Ratio, Volume,} and \textit{Construction Area}. These features demonstrate direct causal effect connections to the target outcome and simultaneously carry causal dependencies with other features within the model. 

Among these three features, \textit{Construction Area} is at most important: It is the only feature that shares a common cause with the outcome (\textit{EUI Heating}), and the common cause being one of the exposure inputs (\textit{Insulation Standard}). This is expected given that the construction area is an input in the EUI estimation. The fact that it shares a cause with the outcome means that blocking the \textit{Construction Area} would close the causal path from: \textit{Insulation Standard $\rightarrow$ Construction Area $\rightarrow$ EUI Heating}, and open a biasing path (a detour connection from exposure to the outcome) as: \textit{Insulation Standard $\rightarrow$ Area $\rightarrow$ Volume $\rightarrow$ EUI Heating }(Figure \ref{fig:4.png}h). This explains the unusual prediction results in Scenario II with variations in Insulation Standard. To correctly estimate the direct effect of \textit{Insulation Standard }on \textit{EUI heating}, we should either involve the feature \textit{Construction Area} in the model to keep the causal path open, or we need to exclude \textit{Construction Area}, \textit{Area}, and \textit{Volume} together to avoid the biasing path. In other words, causal dependencies exist between the building insulation standard, construction area, building area, and volume; controlling the intermediate one and varying the rest leads to a biased sampling situation. 

From an engineering domain perspective, this causal finding conclusion mentioned above is derivable and can withstand cross-validation of domain knowledge, as the construction area serves as a common effect reflecting the configuration of the building area and building insulation standards: It is important to note that a larger building area and volume do not necessarily result in a proportional increase in the construction area. For instance, the thickness of building internal walls (non-loadbearing) and facades within the same insulation standard remains unchanged. Consequently, as the total building area expands, the building construction area proportion correspondingly shrinks. Meanwhile, higher building insulation standards correlate with better thermal isolation behavior for building façades. Better isolation typically equates to a thicker structure installation, hence the increase in construction area. Although we consider the \textit{Construction Area} not directly affecting the \textit{EUI Heating} since we vary the insulation standards, removing this feature from the model means the model samples through possible ranges from training data (refer to Table \ref{tab:tab2}) and hence cancels out the consequential changes of \textit{Insulation Standard}, while building \textit{Area} and \textit{Volume} are fixed, leading to more biases samples.

\subsection{Validation}
Building upon the conclusion from the causal dependencies analysis above, we can state: 

\textit{“To properly investigate the causal effect from the Insulation Standard to EUI, the Construction Area should not be ignored for an unbiased effect estimation.”}

With the same features selected as in Scenario II, \textit{\textbf{Construction Area}} is additionally included. The corresponding performance with the updated feature set is given in Table 4:, while the test case prediction result is illustrated in Figure \ref{fig:4.png}j. Notably, with only a slight decrease in accuracy compared to the performance in Scenario I (Table \ref{tab:tab3}), the prediction trend and uncertainty ranges of the EUI Heating align with the output in Scenario I again.

\begin{table}
	\caption{5-fold cross-validation performance result comparison of different models: Validation Scenario}
\centering
\begin{tblr}{
  width = \linewidth,
  column{1} = {c},
  column{2} = {c},
  hline{1-2,6} = {-}{},
}
\textbf{Model}         & \textbf{R\textsuperscript{2 }} \\
\textit{Decision Tree} & 0.81                           \\
\textit{SVR}           & 0.90                           \\
\textit{ANN}           & 0.96                           \\
\textit{NGBoost}       & 0.90                           
\end{tblr}
\label{tab:tab4}
\end{table}

\subsection{Occam’s Razor for Knowledge Discovery: Identifying the Minimal Sufficient Adjustment Set}
Causal discovery analysis could also contribute to determining the minimal number of required variables thanks to the concept of “minimal sufficient adjustment sets”. A causal DAG helps to answer the following common question in the data-driven process: 

\textit{“Which variables (features) should we include in our model to get an unbiased estimate of the effect?”}

A "minimal sufficient adjustment set" refers to the smallest set of variables that need to be adjusted to reliably estimate a causal effect. These sets can be identified manually \citep{greenland1999causal,shrier2008reducing} or with a computer package \citep{textor2016robust}. In this context, the well-known concept of Occam’s razor is appropriate for the causal model preference \citep{pearl2000models}. 

\begin{figure}[H]
	\centering
	\includegraphics[width=16cm]{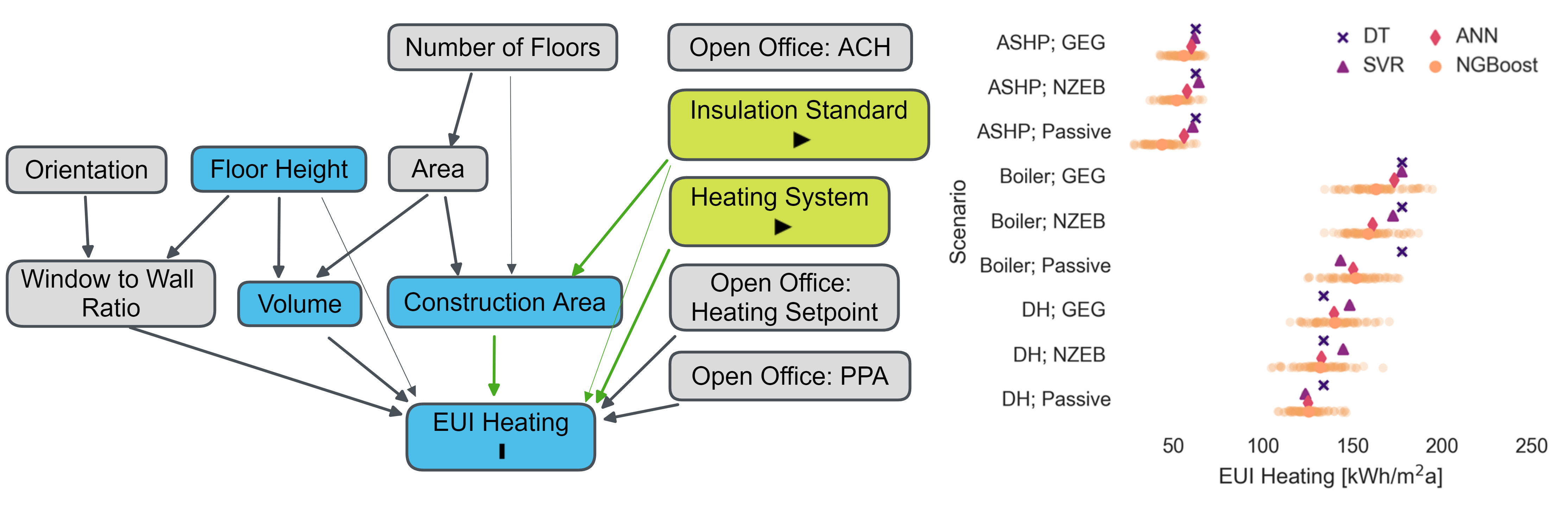}
	\caption{Minimal sufficient adjustment set based on the case: With \textit{Floor Height, Volume }and \textit{Construction Area} as extra inputs, the model generates unbiased estimation with sufficient information from the dataset.}
	\label{fig:5.png}
\end{figure}

Take our case as an example, one minimal sufficient adjustment set would include: \textit{Construction Area}, \textit{Floor Height}, and \textit{Volume}. A skeleton illustration is given in Figure \ref{fig:5.png}. As a result, we observe a similar unbiased trend in the case prediction as in Scenario I (Figure \ref{fig:3.png}a). Combined with the prediction result, we recognize the potential for knowledge discovery in engineering scenarios by interpreting features present in the minimal sufficient adjustment set.

Finally, it is essential to point out that DAGs and the minimal sufficient adjustment set solely provide identification information to ensure unbiased estimation, rather than addressing estimation performance. In engineering contexts, this data-driven process needs to relate to domain knowledge and thus be given context by the task-specific scenario for further analysis.

\section{Discussion}
We utilize a fallout case to demonstrate an easily identifiable error when using data-driven models. However, identifying such errors could be much more challenging for designers in many cases, potentially leading to a distrust in data-driven methods. 

While these easily identifiable errors primarily appear in data-driven methods, similar risks of biased information exist when using first-principles simulations. First-principles simulations, extensively developed by numerous engineers and experts, carry their own biases \citep{rakitta2021cognitive,klotz2011cognitive,zalewski2017cognitive}. The difference is that these biases are often hidden or subtle due to the established and extensively developed nature of these simulations. Cognitive biases \citep{minsky1991logical}, which refer to systematic errors in thinking that affect people's decisions and judgments, can also cause such fallout situations. An example of a cognitive bias relevant in this context is confirmation bias, where engineers might favor information (e.g., a familiar type of design pattern, system deployment, or validation method) that confirms their preexisting beliefs or hypotheses while ignoring or downplaying contrary evidence. This bias leads to a skewed acquisition or utilization of personal domain knowledge. Considering the potential for cognitive biases, simulation results also bear fallout risks and often lack an appropriate adjustment mechanism. In this context, causal analysis serves as a useful tool for identifying potential biases in prior data, thus building a bridge that links and reinforces domain knowledge with data-driven methods. We argue that data-driven methods and first-principles simulations are not inherently conflicting. Rather, combining them may offer a practical solution to manage and mitigate the risk of biased outcomes.

While managing cognitive biases is crucial, another significant aspect to consider is the process of feature selection. In the context of causal analysis, it may seem that the more features (input variables) involved in the modeling process, the more comprehensive the causal skeleton should be. Simply feeding more features into the modeling process doesn’t necessarily contribute to the accuracy improvement. We perceive this as a trade-off between precision and accuracy in describing the case:
\begin{itemize}
    \item More detailed features formalize a good representation of the target case, reducing uncertainty with a more accurate description, but also raise the risk of biased variation analysis.
    \item Using fewer detailed features certainly reduces the risk of biased result analysis; however, a too simple feature representation might overlook important factors that could affect the result and lead to incorrect conclusions.
\end{itemize}

\section{Conclusion}
The evolution of engineering analysis methodologies has fostered synergetic interaction among data, domain knowledge, simulations, and data-driven methods. Our case study highlights the potential pitfalls of relying solely on data-driven methods without incorporating causal analysis. We proved that it is critical to examine causal relationships when performing a data-driven analysis to avoid misleading results. Consequently, we advocate for more attention and involvement in causal inference analysis in the engineering community. Moreover, we believe that extracting invariant and transferable information from data is crucial in bridging the gap between domain knowledge, simulations, and data-driven methods in engineering and transcending individual capabilities’ limitations. 

\section{Acknowledgement}
We gratefully acknowledge the German Research Foundation (DFG) support for funding the project under grant GE 1652/3-2 in the Researcher Unit FOR 2363 and under grant GE 1652/4-1 as a Heisenberg professorship.

\section{Appendix}
\subsection{Mechanism Introduction of Machine Learning Methods}
\textbf{\textit{Tree-based models}} seek to identify optimal split points in the data to enhance prediction accuracy. The term "tree" refers to a decision tree, which forms the foundation of tree-based models. The decision tree algorithm identifies which data feature to split on and when to cease splitting based on information gain criteria (i.e., minimizing entropy in data split). While straightforward to interpret, decision trees are generally weak predictors. Enhanced ensemble methods such as bagging, random forest, boosting \citep{dietterich2000ensemble}, and gradient boosting \citep{natekin2013gradient} have been adapted to improve performance but lead to less interpretable behavior. 

\textbf{\textit{Kernel machines}} utilize a linear classifier to address non-linear problems by defining a separating hyperplane to fit in data and make predictions. A kernel corresponds to a dot product in a typically high-dimensional feature space \citep{hofmann2008kernel}. In this space, estimation methods are linear, and all formulations are made in terms of kernel evaluations, thereby avoiding explicit computation in the high-dimensional feature space. 

\textbf{\textit{Neural networks}} comprise input, hidden, and output layers, where each layer is a group of neurons, loosely modeling the neurons in a biological brain. The connections between neurons (also called nodes) carry associated weights/biases. The data is fed into the network and passes through all neurons with activation functions (which add non-linearity to the output) in the forward propagation to produce output. The backpropagation mechanism \citep{lecun1988theoretical} updates neuron weights/biases according to the difference between prediction and output (loss function evaluation).

\subsection{Modeling Configuration for Generating Training Data }
\begin{figure}[H]
	\centering
	\includegraphics[width=16cm]{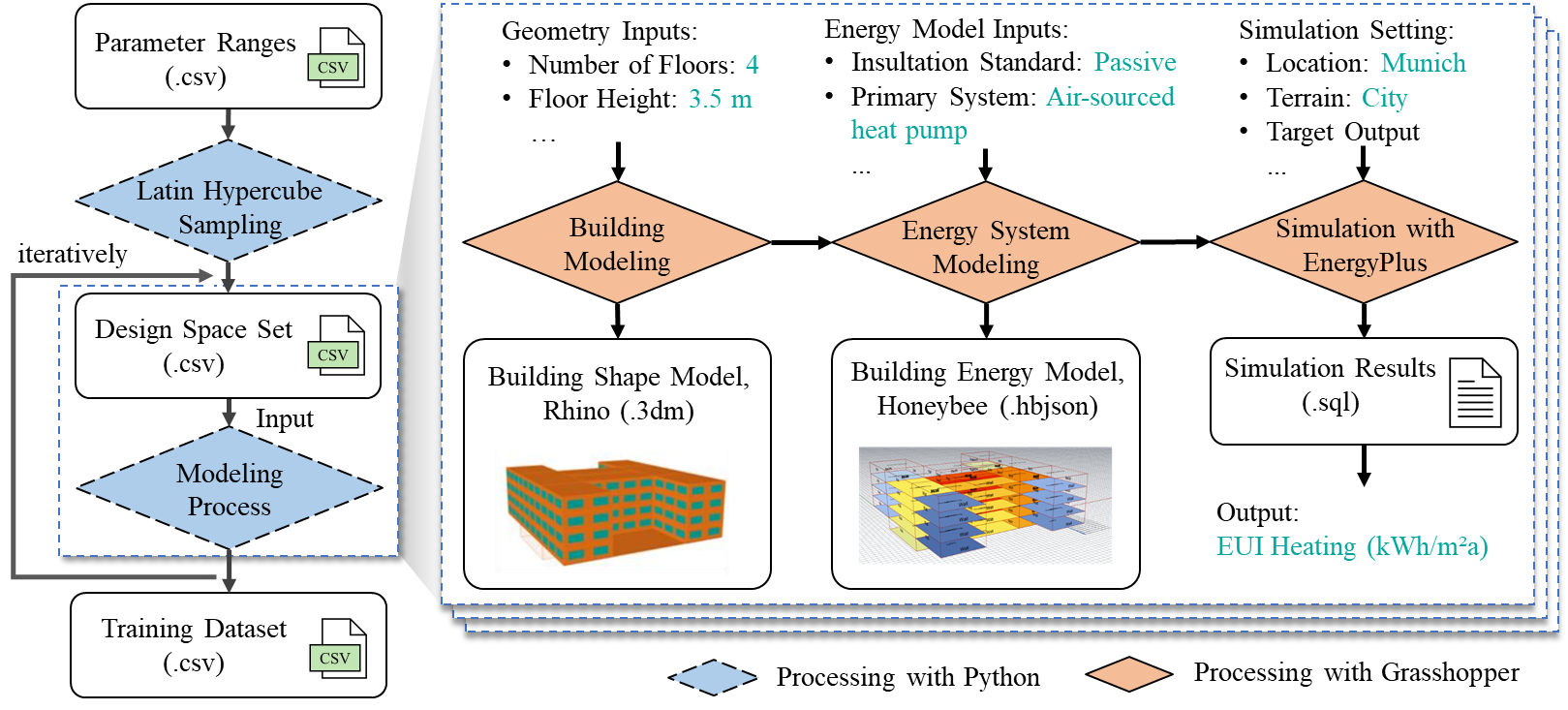}
	\caption{Automatic data generation process with parametric modeling: a generic H-shape office building. The parameter ranges are determined with the consideration of covering the test case scenario and densely sampled with variations. Each sample is fed iteratively into the energy simulation pipeline composited by Grasshopper, Python, and intermediate models. 918 samples were generated as the training dataset.}
	\label{fig:6.png}
\end{figure}

The test case is a mixed-usage 4-floor building named Building.Lab on a tech campus in Regensburg, Germany \citep{Chen2022}. The function of this 1,956 m² building is office and seminar use as well as housing, which consists of four above-ground stories and one underground level with a concrete skeleton structure. For supporting decision-making in energy-efficient building design, we developed a parametric model of an office building in a generic H-shape that covers a wide configuration variety of building components and zones. We varied this model to generate a representative training dataset for well-generalizing models on the target scenarios covering the design space characteristics of the case and similar buildings for performance evaluation. An illustration of the data generation process is given in Figure \ref{fig:6.png}.

For the variation of building insulation standards, we simulated three component thermal characteristic sets based on real-world building energy standards and, from low to high: 2020 German Energy Act for Buildings (GEG), Net Zero Energy Building (NZEB), and Passive House. The standards have different requirements for components’ thermal conductivity (U-values), as presented in Table \ref{tab:tab5}.

\begin{table}
	\caption{Different insulation standard requirements for building component thermal characteristics [W/m²·K]}
\centering
\begin{tblr}{
  width = \linewidth,
  colspec = {Q[362]Q[258]Q[223]Q[125]},
    row{1} = {c},
  column{2} = {c},
  column{3} = {c},
  column{4} = {c},
  hline{1-2,7} = {-}{},
}
\textbf{Insulation standard of U-Values in building components} & \textbf{Base: GEG (2020 German Energy Act for Buildings)} & \textbf{Medium: NZEB (Net Zero Energy Building)} & \textbf{High: Passive House}\\
\textbf{Base plate} & 0.2625 & 0.206 & 0.15\\
\textbf{Roof} & 0.15 & 0.135 & 0.12\\
\textbf{Exterior wall, bearing, above ground} & 0.21 & 0.18 & 0.15\\
\textbf{Exterior wall, bearing, under ground} & 0.2625 & 0.206 & 0.15\\
\textbf{Window} & 0.975 & 0.888 & 0.8
\end{tblr}
\label{tab:tab5}
\end{table}

As for heating systems, three typical building energy systems are simulated: boiler, air-source heat pump (ASHP), and district heating (DH). All systems have been modeled with convective hot water baseboards as their secondary energy system. The hot water loop temperature was 50°C for the air-sourced heat pump system variant and 80°C for the boiler and district heating system variants. The piping system was modeled as adiabatic. The heating setpoint scales a typical office hour schedule to a new target setpoint. During off-work hours (starting from 6 pm), only 75\% of the setpoint is set. Starting at 6 am, setpoints are increased hourly to 85\%, 95\%, and 100\%. The minimum heating temperature is set to 21°C as we referred to the national standard DIN EN 16798-1 \citep{Beuth2022din16798-1}, and we intend to find sustainable and high-performing solutions (all options to be inside category I with PPD<6\%). As the comfort temperature is 22°C ± 2K for environments below 16°C, we chose 21-24°C. In this simulation model, no cooling system and mechanical ventilation were modeled. The zone ventilation was only set by the air change rate per hour based on exterior air volume demands set from DIN EN 16798-1. 

To validate the simulation result, we sampled the generated data (Training data) by different insulation standards and heating systems, as presented in Table \ref{tab:tab6} and Table \ref{tab:tab7}, respectively.

\definecolor{Gray}{rgb}{0.498,0.498,0.498}
\begin{table}[t]
	\caption{Energy Usage Intensity (EUI) Heating distribution, sampled by heating system choice}
\centering
\begin{tblr}{
  width = \linewidth,
  colspec = {Q[446]Q[67]Q[54]Q[67]Q[54]Q[67]Q[54]Q[67]Q[54]},
  row{2} = {c},
  row{3} = {c},
  cell{1}{1} = {r=3}{},
  cell{1}{2} = {c=2}{0.121\linewidth,c},
  cell{1}{4} = {c=2}{0.121\linewidth,c},
  cell{1}{6} = {c=2}{0.121\linewidth,c},
  cell{1}{8} = {c=2}{0.121\linewidth,c},
  hline{1} = {-}{},
  hline{2-3} = {2-9}{Black},
  hline{4} = {-}{Gray},
}
\textbf{Energy Usage Intensity (EUI) Heating [kWh/m\textsuperscript{2}a]} & \textbf{All} &  & \textbf{ASHP} &  & \textbf{Boiler} &  & \textbf{DH} & \\
 & mean & std & mean & std & mean & std & mean & std\\
 & 84.6 & 50.1 & 45.4 & 13.5 & 143.0 & 44.6 & 106.3 & 31.6
\end{tblr}
\label{tab:tab6}
\end{table}

\definecolor{Gray}{rgb}{0.498,0.498,0.498}
\begin{table}[t]
	\caption{Energy Usage Intensity (EUI) Heating distribution, sampled by insulation standard}
\centering
\begin{tblr}{
  width = \linewidth,
  colspec = {Q[446]Q[67]Q[54]Q[67]Q[54]Q[67]Q[54]Q[67]Q[54]},
  row{2} = {c},
  row{3} = {c},
  cell{1}{1} = {r=3}{},
  cell{1}{2} = {c=2}{0.121\linewidth,c},
  cell{1}{4} = {c=2}{0.121\linewidth,c},
  cell{1}{6} = {c=2}{0.121\linewidth,c},
  cell{1}{8} = {c=2}{0.121\linewidth,c},
  hline{1} = {-}{},
  hline{2-3} = {2-9}{Black},
  hline{4} = {-}{Gray},
}
\textbf{Energy Usage Intensity (EUI) Heating [kWh/m\textsuperscript{2}a]} & \textbf{All} &  & \textbf{GEG} &  & \textbf{NZEB} &  & \textbf{Passive} & \\
 & mean & std & mean & std & mean & std & mean & std\\
 & 84.6 & 50.1 & 90.8 & 56.5 & 85.1 & 49.0 & 78.0 & 45.1
\end{tblr}
\label{tab:tab7}
\end{table}

\subsection{Training Process and Result Validation}
During the model training process, a hyperparameter grid-search strategy with 5-fold cross-validation \citep{refaeilzadeh2009cross} is applied for fitting data scheme changes in each scenario for all ML models. From an intuitive understanding, it means the same model with all hyperparameter setting combinations are cross evaluated within the 80/20 split training data, to compare and ensure the models’ best performance for test case validation. The results analysis by three evaluation metrics in all scenarios is presented in Table \ref{tab:tab8}.

\begin{table}[h]
	\caption{5-fold cross-validation performance result comparison of different models, all scenarios.}
\centering
\begin{tblr}{
  width = \linewidth,
  colspec = {Q[185]Q[137]Q[244]Q[88]Q[98]Q[169]},
  row{3} = {c},
  row{4} = {c},
  row{6} = {c},
  row{7} = {c},
  row{9} = {c},
  row{10} = {c},
  cell{1}{2} = {c},
  cell{1}{3} = {c},
  cell{1}{4} = {c},
  cell{1}{5} = {c},
  cell{1}{6} = {c},
  cell{2}{1} = {r=3}{},
  cell{2}{2} = {c},
  cell{2}{3} = {c},
  cell{2}{4} = {c},
  cell{2}{5} = {c},
  cell{2}{6} = {c},
  cell{5}{1} = {r=3}{},
  cell{5}{2} = {c},
  cell{5}{3} = {c},
  cell{5}{4} = {c},
  cell{5}{5} = {c},
  cell{5}{6} = {c},
  cell{8}{1} = {r=3}{},
  cell{8}{2} = {c},
  cell{8}{3} = {c},
  cell{8}{4} = {c},
  cell{8}{5} = {c},
  cell{8}{6} = {c},
  hline{1} = {-}{},
  hline{2,11} = {-}{Black},
  hline{5,8} = {2-6}{Black},
}
~ & ~ & \textbf{Decision Tree} & \textbf{SVR} & \textbf{ANN} & \textbf{NGBoost}\\
\textit{Scenario I} & \textit{NRMSE } & 8.22 & 7.85 & 4.04 & 4.51\\
 & \textit{SMAPE } & 0.15 & 0.14 & 0.10 & 0.09\\
 & \textit{R\textsuperscript{2}} & \textbf{0.86} & \textbf{0.87} & \textbf{0.96} & \textbf{0.95}\\
\textit{Scenario II} & \textit{NRMSE} & 9.70 & 7.81 & 5.35 & 7.48\\
 & \textit{SMAPE} & 0.18 & 0.14 & 0.10 & 0.14\\
 & \textit{R\textsuperscript{2}} & \textbf{0.81} & \textbf{0.87} & \textbf{0.94} & \textbf{0.88}\\
\textit{Validation} & \textit{NRMSE} & 9.58 & 6.81 & 4.43 & 7.05\\
 & \textit{SMAPE} & 0.18 & 0.11 & 0.09 & 0.14\\
 & \textit{R\textsuperscript{2}} & \textbf{0.81} & \textbf{0.90} & \textbf{0.96} & \textbf{0.90}
\end{tblr}
\label{tab:tab8}
\end{table}

\bibliographystyle{unsrtnat}
\bibliography{references}  

\end{document}